\documentclass[twoside]{ae100prg}
\usepackage{graphicx}
\usepackage[breaklinks]{hyperref}
\usepackage{booktabs}

\usepackage{mathrsfs}

\def\espaitemps{({\cal V},g)}
\def\varietat{{\cal V}}
\def\AH{\mbox{A3H}}
\def\B{\mathscr{B}}

\newtheorem{theorem}{Theorem}
\newtheorem{defi}{Definition}
\newtheorem{result}{Result}

\begin{document}
\title{On the stability operator for MOTS and the `core' of Black Holes}

 \author{Jos\'e M. M. Senovilla}
 \address{F\'{\i}sica Te\'orica, Universidad del Pa\'{\i}s Vasco,  Apartado 644,\\
48080 Bilbao, Spain}
 \email{josemm.senovilla@ehu.es}

%
%

\begin{abstract}
Small deformations of marginally (outer) trapped surfaces are considered by using their stability operator. In the case of spherical symmetry, one can use these deformations on any marginally trapped round sphere to prove several interesting results. The concept of `core' of a black hole is introduced: it is a minimal region that one should remove from the spacetime in order to get rid of all possible closed trapped surfaces. In spherical symmetry one can prove that the spherical marginally trapped tube is the boundary of a core. By using a novel formula for the principal eigenvalue of the stability operator, I will argue how to pursue similar results in general black-hole spacetimes.
\end{abstract}

\section{Introduction: basic concepts and notation}

Let $S$ denote a closed marginally outer trapped surface (MOTS) in the spacetime $\espaitemps$.
This means that the outer null expansion vanishes $\theta_{\vec k}=0$, where here
the two future-pointing null vector fields orthogonal to $S$ are denoted by $\vec\ell$ and $\vec k$, the latter is declared to be outer, and we set $\ell^\mu k_{\mu}=-1$ as a convenient normalization.
If in addition the other null expansion is non-positive ($\theta_{\vec\ell} \leq 0$), then $S$ is called a marginally trapped surface (MTS). I will also use the concept of outer trapped surface (OTS) when just $\theta_k <0$ and of future trapped surface (TS) if both expansions are negative: $\theta_k<0$ and $\theta_\ell <0$. A hypersurface foliated by M(O)TS is called a marginally (outer) trapped tube, abbreviated to M(O)TT. For further explanations check \cite{AG,AK,S,S1,Wald}.

\subsection{Stability operator for MOTS}
As proven in \cite{AMS,AMS1}, the variation of the vanishing expansion $\delta_{f\vec n} \theta_{\vec k}$ along any normal direction $f\vec n$ such that $k_\mu n^\mu=1$ reads
\begin{equation}
\delta_{f\vec n} \theta_{\vec k}=-\Delta_{S}f+2s^B\overline\nabla_{B}f+
f\left(K_{S}-s^B s_{B}+\overline\nabla_{B}s^B-\left.G_{\mu\nu}k^\mu \ell^{\nu}\right|_S -\frac{n^\rho n_{\rho}}{2}\,  W\right)
\label{deltatheta}
\end{equation}
where $K_{S}$ is the Gaussian curvature on $S$, $\Delta_{S}$ its Laplacian, $G_{\mu\nu}$ the Einstein tensor, $\overline\nabla$ the covariant derivative on $S$, $s_{B}=k_{\mu}e^\sigma_{B}\nabla_{\sigma}\ell^\rho$ (with $\vec e_{B}$ the tangent vector fields on $S$), and
$$
W\equiv \left.G_{\mu\nu}k^\mu k^{\nu}\right|_S +\sigma^2 \label{W}
$$
with $\sigma^2$ the shear scalar of $\vec k$ at $S$. Obviously $W\geq 0$ whenever $\left.G_{\mu\nu}k^\mu k^{\nu}\right|_S\geq 0$ (for instance if the null convergence condition holds \cite{HE}). Under this hypothesis, $W=0$ can only happen if $\left.G_{\mu\nu}k^\mu k^{\nu}\right|_S=\sigma^2=0$. This leads to Isolated Horizons \cite{AK}, and I shall assume $W>0$ throughout.

Note that the direction $\vec n$ is selected by fixing its norm:
\begin{equation}
\vec n =-\vec\ell +\frac{n_{\mu}n^{\mu}}{2}\vec k  \label{n}
\end{equation}
and observe also that the causal character of $\vec n$ is totally unrestricted.


The righthand side in formula (\ref{deltatheta}) defines a differential operator $L_{\vec n}$ acting (linearly) on the function $f$: $\delta_{f\vec n} \theta_{\vec k}\equiv L_{\vec n} f$.
$L_{\vec n}$ is an elliptic operator on $S$, called \underline{the stability operator} for the MOTS $S$ in the normal direction $\vec n$. $L_{\vec n}$ is not self-adjoint in general, however it has a real principal eigenvalue $\lambda_{\vec n}$, and the corresponding (real) eigenfunction $\phi_{\vec n}$ can be chosen to be positive on $S$ \cite{AMS,AMS1}. The (strict) stability of the MOTS $S$ is ruled by the (positivity) non-negativity of  the principal eigenvalue $\lambda_{\vec n}$ \cite{AMS,AMS1}.

\section{Spherically symmetric spacetimes}
In advanced coordinates, spherically symmetric spacetimes have the line-element
$$
ds^2=-e^{2\alpha}\left(1-\frac{2m}{r}\right)dv^2+2e^\alpha dvdr+r^2d\Omega^2  \, .
$$
where $\alpha$ and $m$ are functions of $v$ and $r$.
For each round sphere defined by $ \{r,v\}=$consts., its future null normals are
$$
\vec\ell= -e^{-\alpha}\partial_r , \hspace{1cm} 
\vec k=\partial_v +\frac{1}{2}\left(1-\frac{2m}{r}\right)e^{\alpha}\partial_r
$$
so that their null expansions are: 
$$
\theta^{sph}_{\vec k} =\frac{e^{\alpha}}{r}\left(1-\frac{2m}{r}\right), \hspace{1cm}
\theta^{sph}_{\vec\ell} =-\frac{2e^{-\alpha}}{r} .
$$
The set $\AH : \hspace{1mm} r-2m(r,v)=0\hspace{2mm} (\Leftrightarrow \theta^{sph}_{\vec k}=0)$ is an  MTT. A3H is actually the only {\em spherically symmetric} MTT : the only spherically symmetric hypersurface foliated by MTSs ---be they round spheres or not \cite{BS}. 

The round spheres are untrapped if $r>2m$, and trapped if $r<2m$. One can further prove \cite{BS} that any closed trapped surface cannot be fully contained in a region with $r\geq 2m$, so that all of them must intersect the region $\{r<2m\}$.
However, how much must a TS penetrate into $\{r<2m\}$?

Let $\varsigma\subset$ A3H be any MT round sphere (i.e., $\theta^{sph}_{\vec k}=0$) defined by $r=r_\varsigma=$const. The variation $\delta_{f \vec n} \theta^{sph}_{\vec k}$ along normal directions simplifies drastically in this case, because $\sigma^2=0$ ($\vec k$ is shear-free ) and $s_B=0$. In other words, most of the terms in the variation formula vanish and the variation simplifies to
$$
\delta_{f\vec n} \theta^{sph}_{\vec k}=-\Delta_{\varsigma}f+f\left(\frac{1}{r_{\varsigma}^2}-G_{\mu\nu}k^\mu \ell^{\nu}-\frac{1}{2} n_\rho n^\rho \, G_{\mu\nu}k^\mu k^\nu\right)
$$
Selecting $f=$constant, the vector $\vec n$ such that the expression enclosed in brackets vanishes produces no variation on $\theta^{sph}_{\vec k}$, meaning that $\vec n$ is tangent to the A3H simply leading to other marginally trapped round spheres on A3H. Let us call such a vector field $\vec m$, so that $\vec m =-\vec\ell +\frac{m_{\mu}m^{\mu}}{2}\vec k $ with
$
\frac{1}{r_{\varsigma}^2}-\left.G_{\mu\nu}k^\mu \ell^{\nu}\right|_\varsigma -
 \left.\frac{m_\rho m^\rho}{2}G_{\mu\nu}k^\mu k^{\nu}\right|_\varsigma =0 \label{m}
$
characterizes A3H. 

Consider now the parts of $\AH$ with $G_{\mu\nu}k^\mu k^{\nu}> 0$ (i.e. $W>0$). 
From the properties of $\vec m$ one deduces that the perturbation along $f \vec n$ will enter into the region with trapped round spheres (that is, $\{r<2m\}$) at points with $f(n_{\mu}n^{\mu} - m_{\mu}m^{\mu}) >0$.
Note that
\begin{equation}
(G_{\rho\sigma}k^\rho k^{\sigma}|_{\varsigma})\, \,  f(n_{\mu}n^{\mu} - m_{\mu}m^{\mu}) =-2(\Delta_{\varsigma}f+\delta_{f\vec n}\theta^{sph}_{\vec k}). \label{deltatheta1}
\end{equation}
In order to construct examples of TSs which lie partly in $\{r>2m\}$, consider the case
$n_{\mu}n^{\mu} - m_{\mu}m^{\mu} >0$.
For this choice the deformed surface enters the region $\{r<2m\}$ at points with $f>0$. Setting $f\equiv a_{0}+\tilde f $ for some as yet undetermined function $\tilde f$ and a constant $a_0$, Eq.(\ref{deltatheta1}) can be split into two parts
\begin{eqnarray*}
(G_{\rho\sigma}k^\rho k^{\sigma}|_{\varsigma})\, a_{0}(n_{\mu}n^{\mu}- m_{\mu}m^{\mu}) +2 \delta_{f\vec n}\theta^{sph}_{\vec k} = 0, \\
\frac{1}{2}(G_{\rho\sigma}k^\rho k^{\sigma}|_{\varsigma})(n_{\mu}n^{\mu} - m_{\mu}m^{\mu}) = -\frac{\Delta_{\varsigma}\tilde f}{\tilde f} >  0 .
\end{eqnarray*}
By our assumptions the first of these implies that $\delta_{f \vec n} \theta^{sph}_{\vec k}  < 0$ if $a_{0}>0$, so that the deformed surface will be trapped. The second, in turn, is a mild restriction on the function $\tilde f$. A simple 
solution is to choose $\tilde f$ to be an eigenfunction of the Laplacian $\Delta_{\varsigma}$, say 
$\tilde f =c_l P_{l}$ for a fixed $l\in \mathbb{N}$ and constant $c_{l}$, where $P_{l}$ are the Legendre polynomials. 

Even more interestingly, we are ready to answer the question of how small the fraction of any closed TS that extends outside $\{r<2m\}$ can be made. The aim is to produce a $C^2$ 
function $\tilde f$ defined on the sphere
(i) obeying the inequality $\displaystyle{-\frac{\Delta_{\varsigma}\tilde f}{\tilde f} >  0}$, and 
(ii) positive only in a region that we can make arbitrarily small. 
By choosing a sufficiently small constant 
$a_0$ requirement (ii) implies that the part of 
the surface extending outside $\{r>2m\}$ can be made arbitrarily small. 
To find $\tilde f$ explicitly, introduce stereographic coordinates 
$\{\rho, \varphi\}$ on the sphere, so that the Laplacian takes the form
$
\Delta_{\varsigma} = \Omega^{-1} \left( \partial_\rho^2 + 
\frac{1}{\rho}\partial_\rho + \frac{1}{\rho^2}\partial_\varphi^2 \right) \ , 
\hspace{2mm} \Omega = \frac{4r_{\varsigma}^2}
{(1+\rho^2)^2} $,
Then, a solution for $\tilde f$ is the axially symmetric function
\begin{equation}
\tilde f (\rho ) = \left\{ \begin{array}{lll} c_1
\left( e^{\frac{1}{2a}(2a-\rho^2)} - 1\right) & & \rho^2 < 4a \\ 
\\ \frac{8c_1a}{e}\frac{1}{\rho^2} -c_1(1+e^{-1}) & & \rho^2 > 4a \ . 
\end{array} \right. 
\end{equation}
This function is $C^2$ (and can be further smoothed if necessary), and it 
is positive only if $\rho^2 < 2a$, that is on a disk surrounding the origin (the pole) 
whose size can be chosen at will. It obeys 
$$
- \frac{\Delta_{\varsigma} \tilde f}{\tilde f} = \left\{ \begin{array}{lll} 
\frac{\Omega^{-1}}{a^2}\frac{2a-\rho^2}
{1-e^{-\frac{1}{2a}(2c-\rho^2)}} & & \rho^2 < 4a \\ 
\\ \frac{32a\Omega^{-1}}{\rho^4}\frac{\rho^2}{(e+1)\rho^2-8a} \ , & & \rho^2 > 4a \ . 
\end{array} \right. 
$$ 
which is always larger than zero. Thus we have proven the following important and perhaps surprising result \cite{BS}.
\begin{theorem}[Bengtsson \& JMMS 2011]
In spherically symmetric spacetimes, there are closed f-trapped surfaces (topological spheres) penetrating both sides of the (non-isolated part of the) apparent 3-horizon $\AH\backslash\AH^{iso}$ {\em with arbitrarily small portions} outside the region $\{r>2m\}$.
\label{th}
\end{theorem}

\section{Cores}
The (future)-trapped region $\mathscr{T}$ of a spacetime
is defined as the set of points $x\in \varietat$  such that $x$ lies on a closed (future) TS \cite{BS}.
This is a space-time concept, not to be confused with the outer trapped region within spacelike hypersurfaces, which is defined as the union of the interiors of all (bounding) OTS in the given hypersurface \cite{AMS,AM}. I denote by $\B$ the boundary of the future trapped region $\mathscr{T}$: $\B \equiv \partial \mathscr{T}$. 

Closed TSs are clairvoyant , highly non-local objects \cite{AK,BS}. They cross MTTs and even enter flat portions of the space-time \cite{ABS,BS0,BS}. In conjunction with the non-uniqueness of MTTs \cite{AG,BS}, this poses a fundamental puzzle for the physics of black holes.
Although several solutions can be pursued, a popular one is trying to define a preferred MTT. Hitherto, though, there has been no good definition for that. We have put forward a novel strategy \cite{BS}. The idea is based on the simple question: 
{\em what part of the spacetime is absolutely indispensable for the existence of the black hole?}

\begin{defi}[Cores of Black Holes]
A region $\mathscr{Z}$ is called the {\em core} of the f-trapped region $\mathscr{T}$ 
if it is a minimal closed connected set that needs to be removed from the spacetime in order to get rid of all closed f-trapped 
surfaces in $\mathscr{T}$, 
and such that any point on the boundary $\partial\mathscr{Z}$ is connected to $\B=\partial \mathscr{T}$ 
in the closure of the remainder.
\end{defi}
\begin{itemize}
\item Here, ``minimal" means that there is no other set $\mathscr{Z}'$ with the same properties and properly contained in $\mathscr{Z}$. 
\item The final technical condition states that the excised space-time $(\varietat\backslash \mathscr{Z},g)$ has the property that $\forall x\in \varietat\backslash \mathscr{Z}\cup \partial \mathscr{Z}$ there is continuous curve $\gamma\subset \varietat\backslash \mathscr{Z}\cup \partial \mathscr{Z}$ joining $x$ and $\B$ ($\gamma$ can have zero length if $\B\cap \partial \mathscr{Z}\neq \emptyset$). The reason why this is needed are explained in \cite{BS}.  
\end{itemize}

In spherically symmetric spacetimes one can prove that
the region $\mathscr{Z}\equiv \{r\leq 2m\}$ is a core \cite{BS}. 
The proof is founded on the previous Theorem \ref{th}.
It should be observed that this is an interesting and maybe deep result, for the concept of core is global and requires full knowledge of the future while A3H is quasi-local. It is thus surprising that $\AH = \partial \mathscr{Z}$.

Actually, one can further prove that in spherically symmetric spacetimes, $\mathscr{Z}=\{r\leq 2m\}$ are the only spherically symmetric cores of $\mathscr{T}$. 
Therefore, $\partial\mathscr{Z}=\AH$ are the only spherically symmetric boundaries of a core.
Nevertheless, there exist non-spherically symmetric cores of the f-trapped region in spherically symmetric spacetimes. This implies the non-uniqueness of cores, and of their boundaries \cite{BS}. Still, the identified core $\mathscr{Z}=\{r\leq 2m\}$ might be unique in the sense that its boundary $\partial\mathscr{Z}=\AH$ is a MTT: we do not know whether other cores share this property or not \cite{BS}. 

To study whether or not Theorem \ref{th} can be generalized to general situations, thereby providing the possibility of selecting a unique MTT as the boundary of a selected core, consider the family of operators, parameterized by a function $z\in C^\infty (S)$, with a similar structure as that of $L_{\vec n}$: $L_{z}f=- \Delta_{S}f+2s^B\overline\nabla_{B}f +  z f$.
Each $L_{z}$ has a principal {\em real} eigenvalue $\lambda_{z}$ ---which depends on $z$--- and the corresponding eigenfunction $\phi_{z}>0$. 
For any given $z$ one easily gets
$$
\oint_S L_z f=\oint_S \left(2s^B\overline\nabla_B f + zf \right)=\oint_S \left(z-2\overline\nabla_B s^B \right)f
$$
in particular for the principal eigenfunction
$$
\lambda_z \oint_S \phi_z =\oint_S  \left(z-2\overline\nabla_B s^B \right) \phi_z \, .
$$
This provides 
\begin{enumerate}
\item a formula for the principal eigenvalue
\begin{equation}
\lambda_z =\frac{\oint_S  \left(z-2\overline\nabla_{B}s^B \right) \phi_z}{\oint_S \phi_z } \, .\label{lambdaz}
\end{equation}
\item bounds for $\lambda_z$
\begin{equation}
\min_{S} \left(z-2\overline\nabla_{B}s^B\right) \leq \lambda_z \leq \max_S  \left(z-2\overline\nabla_{B}s^B \right) \, .
\label{lambdazbounds}
\end{equation}
\item and that
$\lambda_z - \left(z-2\overline\nabla_{B}s^B\right)$ must vanish somewhere on $S$ for all $z$.
\end{enumerate} 

On any MOTS, varying $\theta_{\vec k}=0$ along the direction $\phi_{z}\vec n$ one derives
$$
\frac{L_{\vec n} \phi_z}{\phi_{z }} =\lambda_{z}-z+K_S-s^B s_{B}+\overline\nabla_{B}s^B -\left.G_{\mu\nu}k^\mu \ell^{\nu}\right|_S -\frac{n^\rho n_{\rho}}{2}W \, .
$$
Thus, {\em whenever} $W\neq 0$ on $S$, one can choose for any $z$ a variation vector $\vec m_{z}=-\vec\ell +M_{z}\vec k$ such that the righthand side vanishes
\begin{equation}
M_{z}=\frac{m^\rho_{z}m_{z\rho}}{2}=\frac{1}{W}\left(\lambda_{z}-z+K_{S}-s^B s_{B}+\overline\nabla_{B}s^B -\left.G_{\mu\nu}k^\mu \ell^{\nu}\right|_S\right) \label{M}
\end{equation}
hence $\delta_{\phi_{z}\vec m_{z}}\,   \theta_{\vec k}=0$. 
Observe that this $\vec m_{z}$ depends on the chosen function $z$.
The general variation of $\theta_{\vec k}$ along $\vec m_{z}$ reads
\begin{equation}
\delta_{f\vec m_{z}}\,   \theta_{\vec k}=-\Delta_{S}f+2s^B\overline\nabla_{B}f+f(z-\lambda_{z})=(L_{z}-\lambda_{z})f \label{deltamzeta}
\end{equation}
so that the stability operator $L_{\vec m_{z}}$ of $S$ along $\vec m_{z}$ is simply $L_{z}-\lambda_{z}$ which obviously has a vanishing principal eigenvalue.
The directions $\vec{m}_z$ define locally MOTTs including any given {\em stable} MOTS $S$ \cite{AMS,AMS1}. These MOTTs will generically be different for different $z$. In fact, given that $\forall z_1,z_2\in C^\infty(S)$, $\vec{m}_{z_1}-\vec{m}_{z_2}=\frac{1}{W}\left(\lambda_{z_1}-z_1-\lambda_{z_2}+z_2\right) \vec k$
one can easily prove that
$$
\vec{m}_{z_1}=\vec{m}_{z_2} \Longleftrightarrow z_1 -z_2 =\mbox{const.}
$$
Now, for any given $z$ rewrite $\delta_{f\vec n}\theta_{\vec k} =L_{\vec n} f$ using (\ref{M}) so that
\begin{equation}
\frac{W}{2}f\left(n^\rho n_{\rho}-m^\rho_{z}m_{z\rho}\right)=
(L_{z}-\lambda_{z})f -\delta_{f\vec n} \theta_{\vec k}  \label{n-m}
\end{equation}

Consider the particular function $z=2\overline\nabla_{B}s^B$. 
This may be the natural generalization of the spherically symmetric MTT shown above. Observe that, for such a choice of $z$, and letting $L\equiv L_{2\overline\nabla_{B}s^B}$, its principal eigenvalue (say $\mu$) vanishes, as follows immediately from either (\ref{lambdaz}) or (\ref{lambdazbounds}). Moreover, 
$$
L f =-\Delta_{S}f +2\overline\nabla_{B}(f s^B) =-\overline\nabla_{B}\left(\overline\nabla^B f-2fs^B \right).
$$
so that $L$ is a divergence and thus $\oint_{S}Lf =0, \hspace{3mm} \forall f $.
Moreover, (\ref{n-m}) reduces to
\begin{equation}
\frac{W}{2}f\left(n^\rho n_{\rho}-m^\rho m_{\rho}\right)=
Lf -\delta_{f\vec n}\theta_{\vec k} \label{n-m2}
\end{equation}
where now the vector $\vec m =-\vec \ell +\frac{m^\rho m_{\rho}}{2}\vec k$ is defined by 
$$
\frac{m^\rho m_{\rho}}{2}=\frac{1}{W}\left(K_{S}-\overline\nabla_{B}s^B -s^B s_{B}-\left.G_{\mu\nu}k^\mu \ell^{\nu}\right|_S\right) 
$$
as follows from (\ref{M}). For any other direction $\vec m_z$ defining a local M(O)TT
$$
\frac{W}{2}\left(m_z^\rho m_{z\rho}-m^\rho m_{\rho}\right)=\lambda_z -(z-2\overline\nabla_Bs^B)
$$
and therefore point (iii) above leads to
\begin{result}
The local M(O)TT defined by the direction $\vec m$ is such that any other nearby local M(O)TT must interweave it with non-trivial intersections to both of its sides, that is to say, the vector $\vec{m}_z -\vec m$ changes causal character on any of its M(O)TSs.
\end{result}

Concerning cores, I try to follow the same steps as in spherical symmetry, and thus I start with a function
$f=a_{0}\phi +\tilde f$
for a constant $a_{0}>0$ and $\phi >0$ is the principal eigenfunction of $L$. Then (\ref{n-m2}) becomes
$$
\frac{W}{2}(a_{0}\phi +\tilde f) \left(n^\rho n_{\rho}-m^\rho m_{\rho}\right)=L\tilde f-\delta_{f\vec n}\theta_{\vec k}
$$
that can be split into two parts:
\begin{eqnarray}
\frac{W}{2}a_{0}\phi \left(n^\rho n_{\rho}-m^\rho m_{\rho}\right)=-\delta_{f\vec n}\theta_{\vec k} \label{first}\\
\frac{W}{2}\tilde f \left(n^\rho n_{\rho}-m^\rho m_{\rho}\right)=L\tilde f \label{second}
\end{eqnarray}
Eq.(\ref{first}) tells us that $\delta_{f\vec n}\theta_{\vec k}<0$ whenever $\vec n$ points ``above'' $\vec m$ if $a_0>0$ is chosen.
Therefore, using (\ref{second}) the problem one needs to solve can be reformulated as follows:
{\em 
Is there a function $\tilde f$ on $S$ such that (i) $L \tilde f/\tilde f \geq \epsilon >0$,
(ii) $\tilde f$ changes sign on $S$,
(iii) $\tilde f$ is positive in a region as small as desired?
}
To prove that there are OTSs penetrating both sides of the MOTT it is enough to comply with points (i) and (ii) only. This does happen if $L$ has more real eigenvalues, for any real eigenvalue is strictly positive (as $\mu =0$), hence the corresponding eigenfunction must change sign on $S$, because integration of $L\psi = \lambda \psi$ on $S$ implies $\oint \psi =0$. However, even if there are no other real eigenvalues the result might still hold in general. In any case, the above leads to the analysis of the condition $L\tilde f/\tilde f >0$ for functions $\tilde f$. 

\section*{Acnodledgments}
Supported by grants FIS2010-15492 (MICINN), GIU06/37 (UPV/EHU) and P09-FQM-
4496 (J. Andaluc\'{\i}a--FEDER) and UFI 11/55 (UPV/EHU).

\section*{References}
\bibliography{PragaProc}

\end{document}